\documentclass[a4paper,12pt]{article}

\usepackage{fullpage} %
\usepackage{times}
\usepackage[left=2.5cm,right=2.5cm,top=2.5cm,bottom=2.5cm]{geometry} %

\usepackage{setspace}

\usepackage[show]{todo}
\usepackage{graphicx}        %

\usepackage[numbers,sort]{natbib}

\usepackage{hyperref}

\usepackage{authblk} %

\newcommand{\secRef}[1]{Section~\ref{sec:#1}}

\begin{document}

\title{An interview based study of pioneering experiences in teaching and learning Complex Systems in Higher Education}

\author[1,$\dagger$]{Joseph T. Lizier}
\author[1]{Michael S. Harr\'e}
\author[2,3]{Melanie Mitchell}
\author[3,4,5]{Simon DeDeo}
\author[1,6]{Conor Finn}
\author[7]{Kristian Lindgren}
\author[8]{Amanda L. Lizier}
\author[9]{Hiroki Sayama}

\affil[1]{Centre for Complex Systems and Complex Systems Research Group, Faculty of Engineering and IT, The University of Sydney, NSW, Australia}
\affil[$\dagger$]{Corresponding author: \href{joseph.lizier@sydney.edu.au}{joseph.lizier@sydney.edu.au}}
\affil[2]{Department of Computer Science, Portland State University, Portland, OR, USA}
\affil[3]{Santa Fe Institute, Santa Fe, NM, USA}
\affil[4]{School of Informatics and Computing, Indiana University, Bloomington, IN, USA}
\affil[5]{Department of Social and Decision Sciences, Carnegie Mellon University, Pittsburgh, PA, USA}
\affil[6]{CSIRO Data 61, Marsfield, NSW, Australia}
\affil[7]{Department of Space, Earth and Environment, Chalmers University of Technology, G\"oteborg, Sweden}
\affil[8]{School of Education, University of Technology Sydney, NSW, Australia}
\affil[9]{Center for Collective Dynamics of Complex Systems and Department of Systems Science and Industrial Engineering, Binghamton University, State University of New York, NY, USA}

\maketitle

\begin{abstract}
Due to the interdisciplinary nature of complex systems as a field, students studying complex systems at University level have diverse disciplinary backgrounds. This brings challenges (e.g. wide range of computer programming skills) but also opportunities (e.g. facilitating interdisciplinary interactions and projects) for the classroom. However, there is little published regarding how these challenges and opportunities are handled in teaching and learning Complex Systems as an explicit subject in higher education, and how this differs in comparison to other subject areas.
We seek to explore these particular challenges and opportunities via an interview-based study of pioneering teachers and learners (conducted amongst the authors) regarding their experiences. We compare and contrast those experiences, and analyse them with respect to the educational literature. Our discussions explored: approaches to curriculum design, how theories/models/frameworks of teaching and learning informed decisions and experience, how diversity in student backgrounds was addressed, and assessment task design. We found a striking level of commonality in the issues expressed as well as the strategies to handle them, for example a significant focus on problem-based learning, and the use of major student-led creative projects for both achieving and assessing learning outcomes.
\end{abstract}

\newpage
\section{Introduction}

Complex systems is a quantitative interdisciplinary science dealing with concepts such as collective behaviour, emergence and self-organisation, using tools including systems thinking, agent-based modelling (ABM), complex networks, game theory and information theory \citep{mitch09,sayama15}.
Formally its roots lie in areas such as chaos theory and nonlinear dynamics, cybernetics, systems theory dating back to the mid-20th century, however it only became formalised as a nominal field in the 1980s.
The first explicit courses in Complex Systems at the Higher Education level soon followed in the 1980s and 1990s, and in particular the first Masters in Complex Systems was launched by Chalmers University of Technology in 2000.

Due to the interdisciplinary nature and reach of complex systems as a field, students in these courses have diverse backgrounds across physics, mathematics, computer science, engineering, biology, neuroscience, economics and other fields. This brings challenges (e.g. diversity of skills, computer programming and analysis ability) but also opportunities (e.g. facilitating interdisciplinary interactions and projects, and authentic applications).

The literature regarding teaching and learning complex systems primarily considers teaching such principles in general and/or teaching primary and high school students \citep{jac06a,hmelo00,hmelo15,yoon17}, embedding such principles within other courses \citep{kellam07} or cataloguing what is being taught in Higher Education courses \citep{sayama17}.
Here however, not only do we focus on teaching and learning complex systems as an \textit{explicit subject} in Higher Education (often in dedicated degree programs) in component \textit{course-work based} units of study (UoS)\footnote{Teaching these principles in supervised capstone and research project based units of study is considered out of scope here.},
we seek to explore the \textit{experience} of how the aforementioned challenges and opportunities are handled in teaching complex systems.
Little literature is available directly on such experiences, and the primary body of knowledge here currently resides in the experience of pioneering teachers and learners in this space.
While there are examples of reflections on individual courses in Higher Education, e.g. \citet{porter18} reflects using the self-lens \citep{brook95} on teaching networks at Master's and more senior undergraduate level, we seek a more broad comparison between multiple teachers and across a variety of areas within complex systems.
As such we have explored the challenges and opportunities in teaching and learning in complex systems in coursework modules via an interview-based study with several such subjects (amongst the authors) on their experiences. We compare and contrast those experiences, and discuss them with reference to the education literature.

This paper begins by describing the methodology and introducing the participants, providing background on their experiences and why they were selected.
The remainder of the paper then reviews the different perspectives encountered regarding each focus area in turn, being: approach to curriculum design, how theories/models/frameworks of teaching and learning informed teacher's decisions and experience, how student diversity was dealt with, and assessment task design.

\section{Method}

The concept for the study was formed in the context of planning for a new Master of Complex Systems degree being launched at The University of Sydney (USyd) in 2017, into which Dr. Joseph Lizier (JL) and Dr. Michael Harr\'e (MH) are teaching.
Dr. Lizier and Dr. Harr\'e co-teach (half of) CSYS5010 ``Introduction to Complex Systems'', as well as co-teaching CSYS5030 ``Self-organisation and criticality'' including modules on information theory (JL) and critical behaviour (MH).

Dr. Lizier sought to gain insights into how experiences of teaching and learning complex systems had guided teachers at other institutions, as input for the design work for these new courses, and so held interviews with several teachers in his network.
The participants were identified for the study due to their established profiles as complex systems teachers, as well as similarity in the subject areas being taught to those of Dr. Lizier.

The interviews were semi-structured, with four common focus areas (approach to curriculum design, how theories/models/frameworks of teaching and learning informed teacher's decisions and experience, how student diversity was dealt with, and assessment task design).
Thematic analysis was performed after the interviews to compare and contrast the perspectives seen within each area. These perspectives are discussed with reference to the education literature in order to tie the practice of teaching and learning complex systems to established education theory.

Background on each participant in the study is as follows.
The remainder of the paper then discusses the perspectives on each focus area in turn.

Prof. Melanie Mitchell (MM) from Portland State University (PSU) and Santa Fe Institute (SFI) %
has recently been teaching CS 346U ``Exploring Complexity in Science and Technology'' at PSU (3rd year computer science undergraduate course).
Prof. Mitchell also ran the SFI Complex Systems Summer School for several years, was the founding Director of the \href{https://www.complexityexplorer.org/}{Complexity Explorer} (MOOC) project at SFI, and designed and delivered their ``Introduction to Complex Systems'' course. %
Additionally, Prof. Mitchell has recently been consulting with Arizona State University regarding curriculum development for their new Master of Complex Systems program.

The interview with Assistant Prof. Simon DeDeo (SDD) from Indiana University (IU)\footnote{Now at Carnegie Mellon University.} and SFI focussed on his recent teaching of I400/590 ``Large-Scale Social Phenomena'' at IU (mixed undergraduate/postgraduate course in cognitive science as well as informatics and computing programs), as well as the ``Maximum Entropy methods'' tutorial he designed and delivered for Complexity Explorer.

Prof. Kristian Lindgren (KL) was one of the architects of the Master of Science in Complex Adaptive Systems at Chalmers University of Technology, which in 2000 was the first Masters program in complex systems offered in the world.
This interview focussed on the design of the degree itself, as well as Prof. Lindgren's teaching of the ``Information Theory of Complex Systems'' (which has run for over 20 years) and (the more recently developed) ``Game Theory and Rationality'' elective units for this degree.

Prof. Hiroki Sayama (HS) has been teaching several courses into Bioengineering and more recently Systems Science programs at Binghampton University.
These include: BE-201/BIOL-333 Self-Organizing Systems (2nd/3rd year undergraduates), SSIE-523 Collective Dynamics of Complex Systems (graduates and 4th year undergraduates), SSIE-641 Advanced Topics in Network Science (advanced graduates) and SSIE-500 Computational Tools (graduates).

Finally, we included an interview with Mr. Conor Finn (CF), who attained a Master of Science in Complexity Science from the University of Warwick between 2013 and 2014.
The goal of this interview is to allow us to view the experience of learning in this area through the student lens \citep{brook95}, in order to complement the experience of the teachers.

\section{Approach to curriculum design}
\label{sec:curriculum}

Curriculum design was widely acknowledged as presenting a challenge, for reasons including the diversity of student backgrounds (dealt with primarily in \secRef{diversity}) and the fact that the field is relatively new.
On the latter point, Prof. DeDeo surmised: ``\textit{Physics has centuries of how the pedagogy is put together: we don't!}''
Indeed, \citet{ashwin14a} describes the evolution of curricula in terms of moving from knowledge-as-research to knowledge-as-curriculum and knowledge-as-student-understanding, and from this perspective it is clear that curricula in our field is at an early stage of its evolution, with a heavy influence of recent research findings.

Furthermore, Prof. Mitchell observed that figures in our field have very different opinions about which topics are essential for complex systems courses in higher education, and this reflects 
\citet{ashwin14a} stating (citing \citet{bern00}) that this evolution is often contested as ``\textit{different voices seek to impose particular versions}''.
One example raised here by Mr. Finn is the minimal attention given to ABM in the Warwick program, relative to the seemingly central position it is given in other programs. Mr. Finn suggested this could be due to the positioning of the Warwick program from a more mathematical than computing perspective, focussing instead on more mathematical approaches such as Markov modelling.
Indeed, the influence of the teacher's previous background and experience on their curriculum design was clear to themselves and others, with for example Prof. Mitchell constructing her introductory course around the topics previously explored in her introductory book \citep{mitch09}, and Prof. Lindgren describing the initial construction of the Chalmer's Master program around the areas of interest of the designers.
Similarly, information theory was cited in a number of interviews as being either included (KL,SDD,JL) or excluded (HS) due in large part to the background of the teacher.

Certainly this may feel challenging to the teachers involved, and some self-doubts about getting the curriculum ``right'' linger, yet this does appear to be a natural and important part of curriculum evolution and contestation raised by \citet{ashwin14a}.
Indeed, because of the early stage of curriculum development in our field, many of the pioneering teachers here are influential researchers in the field.
This is not only a positive because of the general importance of the overlap between teaching and research, but as argued by \citet{brew12a} this tight knit between research and teaching in the one ``community of practice'' more naturally leads to Conceptual Change Student-Focussed (CCSF) teaching styles which encourages active learning (discussed further in \secRef{scholarly}) and deeper learning outcomes.
Crucially, \citet{brew12a} interprets such styles as more readily aligned with the collaborative building of knowledge rather than viewing knowledge as objective.
As such, the perspective provided by the literature allays concerns regarding getting the content ``right'' to a large degree, and indeed reveals the strong integration of our backgrounds and research with our teaching as a strength rather than a limitation.

Moving on to the process of curriculum formulation itself, while none of the interviewees explicitly identified a theoretical framework that was followed, key points of best practice are clearly identifiable.
In particular, principles of constructive alignment \citep{biggs11} between learning outcomes, activities and assessments were clear across the interviews.
To begin with, in each interview evidence of deep reflection on the development of learning outcomes emerged without prompting, as well as how these had been tailored to the teaching context.
A key example here was Prof. Lindgren's description of the overarching learning outcome for their Masters' program as a whole, being the goal for students to come away with ``\textit{a toolbox that is useful to implement and analyse models}'' of complex systems, which naturally prompts alignment through related activities and assessment.
Furthermore, learning outcomes were evident to students as well. Mr. Finn described that the key learning outcome he saw in his degree was attaining a new way of thinking about complex systems, and also described (the alignment of activities via) the understanding that implementation of models was a means to an end with the end being the learning outcome of what the model showed.
This perspective resonated amongst the interviewees (and will be revisited in later sections), and exemplifies how making such learning outcomes so explicit can lead to deeper learning and engagement in this field.
Additionally, the role of assessments as an important tool to \textit{achieve} (in addition to testing) the learning outcomes was also clear, in alignment with CCSF strategies as described by \citet{trig13a}, which we will continue discussion on in \secRef{assessment}.

Finally, another common experience of the interviewees was an initially perceived scarcity of teaching resources because the field is new, with a common response of significant work being undertaken by the interviewees to design and develop what have become widely renown resources.
For example, Prof. Mitchell and Prof. Sayama have both written introductory complex systems books \citep{mitch09,sayama15}\footnote{Though in one case (MM) the book prompted development of the course, while in the other (HS) the book was developed in response to low ratings from students on the previous book.}, whilst the academics at Warwick also developed a book \citep{ball13} from their own lecture notes in a similar fashion to \citet{porter18}.
These activities also include (open source) software toolkits, with Prof. Sayama developing PyCX \citep{sayama-pycx} for simple complex systems, networks and ABM simulations, and Dr. Lizier developing JIDT \citep{liz14c} for information-theoretic analysis of complex systems including a graphical user interface (GUI) for making calculations without writing code.
Prof. Mitchell has also developed video lectures used in both the ComplexityExplorer MOOCs and her teaching at PSU.
The ComplexityExplorer videos also include guest speakers and interviews (and are freely available online for other teachers to utilise).

\section{Dealing with diversity of student backgrounds}
\label{sec:diversity}

The major theme -- constant across all interviews and echoed by \citet{porter18} -- was that the key challenge in teaching and learning of complex systems at the tertiary level is the diversity of student backgrounds.
The issue is caused by the wide range of domains that complex systems science is applicable to -- the generally expressed desire for our teaching to be accessible across all of these disciplines leads to these courses often being open to students from all faculties and/or backgrounds.
The impact of this diversity though is a wide variation in mathematical analysis and computer programming capabilities amongst students, both of which are central to expert application of the toolset being taught.
Put simply by Prof. Lindgren: ``\textit{If you cannot do maths or computing, you struggle}''.

On one hand, there is a simple solution to this issue: Chalmers University moved to raise their entry requirements in terms of mathematical and computational capabilities, having found that ``\textit{it was difficult for the program to be comprehensive if it was opened up to all backgrounds}'' (KL).
They have found this solution to suit their context of a program situated within a Physics school, and with a critical mass of capable students sustaining the program.
Yet this solution does not suit all contexts, and indeed 
there are also very good pedagogical arguments for running a more inclusive program.
My interviewees consistently highlighted the unique learning opportunity afforded by having such a diversity of students together in one classroom:
e.g. ``\textit{putting students from different backgrounds together is not an experience they often have}'' (MM); ``\textit{having people with different experiences working together was sometimes very good, sometimes not, but either way was a very nice experience to have in the classroom}'' (as a student -- CF).
This is valuable as such interactions are highly authentic to the practise of complex systems in addressing real problems, which not only provides crucial learning experiences but is also important for student motivation \citep{ashrowe14}.

In accepting such diversity, the question then becomes how to address it in our teaching.
To some extent, this must involve simplifying the curriculum, with for example much mathematical content left out in some cases (MM and SDD) in order to make the content accessible.
Similarly, Prof. Sayama moved to rebalance his teaching of dynamical systems towards hands-on modelling activities (e.g. modelling and stability analysis) rather than chaos and bifurcations directly in order to bring the concepts to life.
Then, creative options abound for enabling computer programming and modelling in more simple ways.
The program at Warwick runs an intensive introduction to programming in Matlab course for a week before the degree program begins.
More simply, the \href{http://ccl.northwestern.edu/netlogo/}{NetLogo} \citep{wil99a,tis04a} platform for ABM has a reputation in our field as easy to use for beginners, offering graphical-user interfaces (GUIs) to facilitate simple interactions and designs, coupled with lower level software design interfaces to provide more powerful features to advanced users.
Prof. Mitchell, Dr. Lizier and Dr. Harr\'e have been using this as the primary teaching platform for their introductory courses and reported it to be particularly suitable.
Prof. Sayama also cites the need for a ``\textit{common ground that people could learn quickly and start building complex systems}'' with, but developed his own toolkit, PyCX \citep{sayama-pycx}, as an alternative to NetLogo in order to facilitate students learning the more widely-usable Python language.
In addition, Prof. Mitchell described her use of different levels of formative assessment tasks in order to more effectively help students with lower computational capabilities, whilst also facilitating deeper tasks for more advanced students.
The value here may be interpreted through the ``zone of proximal development'' concept \citep{vyg86}, in that more effective learning outcomes are facilitated by tuning the challenge in the tasks to levels appropriate to stretch each student.

Prof. DeDeo is taking the simple computational tools approach further still by currently exploring the extent to which he is able to teach the required computational analysis using spreadsheets. This may sound controversially simple to computing scientists, however almost all students are already familiar and comfortable with this environment, and analysis can be performed with little computational skill in an authentically professional manner.
Dr. Lizier has taken a similar approach in adapting his information-theoretic analysis toolkit JIDT \citep{liz14c} to facilitate simple analysis by students via a point and click GUI interface, which does not require code to be written.
Moreover, the JIDT GUI also generates code templates that more capable students can extend in order to perform more complicated analysis.

Crucially, learning outcomes were consistently stated to be focussed on understanding and applying key principles rather than technical details regarding mathematical analysis and pro\-gramm\-ing-based construction of models.
As summarised by Prof. DeDeo: ``\textit{I want to teach the concepts, not programming}''.
This was the case even in more technical courses (KL and CF), where construction of an intricate computational model may have been a necessary step but it was the insights produced using the model provided that were assessed. %
Similarly, \citet{porter18} states that his learning outcomes were not about whether students can program but ``\textit{rather that they can successfully use, understand, and interpret the output of computations}''.
This provides impetus for the use of simplified computational environments such as NetLogo and spreadsheets, allowing focus on addressing higher-order cognitive process dimensions (c.f. Bloom's revised taxonomy \citep{krath02}) such as creating and critical evaluation.
Indeed, this is quite a positive in either the more focussed or inclusive contexts, aligning well with Conceptual-change student-focussed (CCSF) teaching styles that support deeper learning \citep{rams03}. %

Additionally, the mix of abilities of students in the class affords the opportunity to engage stronger students in becoming ``informal tutors'' helping those weaker.
This was observed in all interviews and in \citep{porter18}, in particular regarding mathematical and software tasks. %
From the student perspective (CF) this was perceived to be quite valuable.
Indeed, the program at Warwick was noted as fostering an environment encouraging the students to work together in this manner, via informal tutorials with problem solving left mainly for the students to guide as a group.
Aspects seen as crucial for the success of this approach were the small size of the cohort, but also that a sense of identity as cohort and group culture was actively fostered in this Masters degree program.\footnote{We will revisit this technique in \secRef{scholarly}.}
Importantly, these interactions between students were thought to also help the more advanced students as well via gaining a deeper understanding of the material.
Such interactions are widely discussed in the social constructivist literature \citep{harland03,vyg86,wass14}.
A social constructivist approach requires reframing the role of the teacher as a facilitator rather than content deliverer, providing scaffolding for group activities in which students are empowered to take on the role of a ``more capable peer'' insofar as their current skill level allows \citep{harland03}.

And finally, Prof. Mitchell and Prof. Sayama raised the important point in this context that students who don't bring mathematical or software skills to the class do bring other expertise (e.g. biological knowledge) which should be usefully shared with the class, in particular in solving problems from those domains, and indeed this is important to bring out such that these students also feel valued in the classroom.
Furthermore, Prof. Sayama noted that typically ``\textit{technically-skilled students trust their technical skills too much}'' and need to learn from these domain experts who in turn have the knowledge but need to learn how to formulate it.

Several other strategies were raised which were seen to help address the diversity of backgrounds; we will discuss these in \secRef{scholarly} regarding the scholarly basis of teaching here.

\section{Scholarly basis of teaching}
\label{sec:scholarly}

When asked about how their teaching was informed by the educational literature, the importance of problem-based learning \citep{savery06} was consistently identified by the interviewees as central to the scholarly basis of teaching complex systems.
This was perhaps best summed up by Prof. Lindgren who stated that in this field undertaking the whole iterative process of ``\textit{model development, implementation, analysis, feedback}'' etc., is \textit{necessary} to attain a deep understanding. That such a problem-based form of learning was particularly suitable for our field was contrasted with mathematics where equations and direct effect of parameters are a focus, or computer science where focus often stops at model implementation.
This aligns with findings on the importance of active learning for complex systems in other contexts, e.g. the study of \citet{hmelo00} regarding teaching concepts to primary school children, and a number of design principles for learning complex systems posed by \citet{jac06a} including ``Experiencing Complex Systems Phenomena'' and ``Constructing Theories, Models, and Experiments''.
And of course this aligns with the importance of experimentation for learning in a wider sense, e.g. via Kolb's theory of experiential learning \citep{kolb74}.

Using a problem-focus in teaching was highlighted as a strength of the Warwick approach by Mr. Finn, who also observed that this reflected their approach to research.
Additionally, Prof. DeDeo had found improved student engagement using the pedagogical technique of starting from an example and then bringing theory in, but also takes this further in actively threading that example throughout the class and getting the students to experiment with and apply their analysis.
For example, as an early introduction to the concepts of uncertainty and entropy in information theory, he has the class play pairwise games of rock-paper-scissors, record their student-generated data and then has them apply the techniques presented in class to analyse that data.
Dr. Lizier has used a similar gamification of the concepts of uncertainty and entropy using the ``Guess Who?'' board game.
\citet{porter18} asks students to take pictures of a local network, and to identify nodes and edges and other features.
Prof. DeDeo summarised the impact of these techniques: ``\textit{As soon as you get them to do something, it changes the whole dynamic \ldots The more that I got students to generate data, and in theoretical cases the more I got them to write ideas up on the board, the more engaged they were, the more they were able to  \ldots use the concepts well.}''
Similarly, in the words of Prof. Sayama: ``\textit{Students are quite excited to be doing things quickly: when they see ABMs moving before their eyes it's a real `aha' moment.}''.

Furthermore, Prof. Mitchell tried flipped classroom techniques \citep{lage00} in the most recent semester, since this meant the classroom was entirely devoted to activities (centred around experimentation with ABM simulations).
Importantly, her students liked the approach, and it led to better learning outcomes.
Her students reported fondness for the video lectures, in particular the ability to speed them up and re-watch them. As such, the use of videos in this blended fashion is an important tool in addressing the diversity of capabilities in the classroom (as identified in \secRef{diversity}).
In a similar way, Prof. DeDeo has left as much mathematics out of his lectures as possible, leaving these for individual reading since students absorb it at different rates.

Clearly the aforementioned approaches are very good teaching practice \textit{in general}, and problem-based learning is known to have good learning outcomes for computational skills \citep{nuu05} and to be particularly suitable for interdisciplinary learning \citep{stentoft17}.
This prompts questioning of the extent to which these approaches are helping teaching and learning \textit{specifically} for complex systems rather than in general,
and we argue that the approaches are specifically a good fit with complex systems to a large degree.
We see this in the findings of \citet{hmelo00} and \citet{jac06a} above regarding teaching complex systems in different contexts.
Furthermore, the problem focus aligns particularly well with complex systems, given its foundation in attempting to identify common universal concepts and approaches that can address fundamentally similar \textit{problems} across various systems and fields.
Indeed, we must note that complex systems has a teaching advantage over other fields because (HS) ``\textit{it is simulation based}'' and so ``\textit{students can get immediate responses and feedback}''.
Keeping such problems as a focus also helps to address the diversity of students in the classroom, first by engaging and motivating them and providing authenticity to their backgrounds \citep{ashrowe14} but also because, as identified by Prof. DeDeo, it gives the students a common footing to start from, and solving the problem empowers them to feel like they could solve other problems in a similar way.
Indeed, Prof. DeDeo (echoing \citet{sternberg08}) highlights that such an approach -- beginning ``\textit{with the concrete and then sticking the abstract on}'', informed from the cognitive sciences -- is the reverse of dominant paradigms in teaching quantitative sciences where the method is generally the goal and problems are a means of demonstrating them.
This resonates with the importance placed on performance in an interdisciplinary context by \citet{boix07a},\footnote{This was particularly regarding assessment, as discussed in \secRef{assessment}.} emphasising that knowledge is not properly understood in this context until it is used, and that while integration of knowledge, methods etc. is what is being learned it is a ``\textit{means to a purpose}'' (solving a problem), ``\textit{not an end in itself}''.

As a counterpoint, Prof. Sayama noted a distinction in this area between more junior undergrad and more senior undergrad or graduate students. He made a conscious decision not to include as much in the way of examples for the more junior students so that they did not misunderstand the learning outcomes, whereas graduate students are more mature and can understand the difference between the theory and examples.
He suggested this may be another reason underlying why most courses in Complex Systems are at the graduate level.

Taking problem-based learning further again, the interviews revealed a number of instances of classes being facilitated so that students take responsibility for and to some extent direct their own learning, very much in alignment with a CCSF style of teaching \citep{rams03} and Vygotsky's social constructivist theories of learning \citep{vyg86}.
Mr. Finn described the way in which tutorials in the Warwick program were run in a very open-ended, informal fashion, with model-solutions generally not given. The problems were left for students to solve as a group, in the context of much encouragement to play. This was reported to have worked particularly well and Mr. Finn found it to be rewarding. It is likely that the small group cohort may have been a contributing factor in the success of this approach.
Dr. Lizier and Dr. Harr\'e routinely ask students to identify the use of the tools under discussion in their own background domains, and make brief presentations on these examples to the class; this helps to address the diversity of backgrounds of students and to bring the theory to life in areas that they are interested in.
More deeply, Prof. Lindgren has coordinated a unit of study solely composed of student-led seminars on research topics in complex systems, where students take turns to choose a topic, arrange discussion seminars, select learning materials and exercises / activities. 
Prof. Lindgren reports a high level of student engagement and learning outcomes, and the success of this approach has been reflected in the student evaluations and large increases in enrolments for the coming semester.

Finally, the impact of external influences on teaching and learning was noted during the interviews. %
For example, Mr. Finn described the teaching and learning space at Warwick, being a dedicated area combining lecture theatre, tutorial room and kitchen area, all in the middle of the research centre. This encouraged the students to linger and socialise, and fostered interaction with faculty members and postgraduate research students, not just during the Centre's forums but also over lunch and transient interactions.
Indeed, the role of spaces in social learning has been widely studied, e.g. \citep{solomon03}, and here from the student perspective this was seen as particularly important given the disparity of backgrounds that the group started with, typical in complex systems.

\section{Assessment task design}
\label{sec:assessment}

The principle observation to emerge regarding approaches to assessment was that major student-driven creative projects were seen in coursework units across all interviews with teachers and in the report of \citet{porter18}.
These were not necessarily replacing an end of semester exam or homework tasks, which were included in various ways in these units of study, nor necessarily a group project, though these were often the case.
These projects stood out however because in all cases they were reported as very successful and popular with students\footnote{As evidenced by formal and informal student feedback, enrolments, and teaching awards in several cases, e.g. \citep{porter18}.}, and there was a striking level of similarity between them.
The projects were student-driven in that the students were required to identify their own problem or data set to focus on, then implement a model and/or analysis to investigate that using the tools that had been learned during the semester.
The model/analysis methods here varied across agent-based modelling (MM, HS, JL/MH), information theory (SDD, JL) and game theory (KL), as well as networks in the report of \citet{porter18}\footnote{The projects in \citep{porter18} appear slightly more constrained to one sub-area of networks, which the teacher changes each year.}.
The assessments were staged, always beginning with some type of proposal or preliminary report, often including an oral presentation of results (preliminary or final) after the implementation, and always with a final report. 

Perhaps such similarity should not be so surprising; after all, as described by \citet{james14a} assessment design takes place within a `learning culture' which is formed in part by the academic discipline.
Surprising or not, this prompts the question of whether the designs here are simply following modern best practice in general -- they are certainly rather common in computer science --  or are they specifically suitable for assessment in complex systems?

Examining these designs in the context of the literature suggests many explanations for why they work well in general.
For example, the design clearly reaches the higher order cognitive process dimensions of ``Create'' and ``Analyze'' in Bloom's revised taxonomy \citep{krath02}, which enables deeper learning approaches and engagement \citep{rams03}.
Also, the importance of challenges and problem-solving for learning -- which would be facilitated by such projects -- is well-recognised \citep{boud12a}.
Furthermore, the use of staged assessment is best-practice in allowing feedback of instructor comments to cycle forward into the next assessment stage \citep{beaumont11}.
Beyond these however lie more specific reasons why these assessments work well in the complex systems domain.

The projects were identified as particularly authentic to real ways of working in complex systems, c.f. \citep{ashrowe14}, being a natural culmination of the ``\textit{model development, implementation, analysis, feedback}'' process described by Prof. Lindgren in \secRef{scholarly}.
The importance of assessing the actual performance of students using such processes rather than simply understanding them or performing smaller parts of them is highlighted in an interdisciplinary context by \citet{boix07a}.
Prof. Mitchell describes the use of such projects as facilitating ``\textit{real complex systems work happening inside the classroom}'',
Prof. Lindgren suggested that the students learn a lot about real project work in this setting, and \citet{porter18} suggests that the use of such projects prompted changes in other aspects of his course to make them ``\textit{`more realistic' with respect to what practitioners in network science do}''.
Indeed, the use of project-based assessment to build competence in using such processes reflects current discussions in the higher education literature around sustainable assessment \citep{boud00,boud16,fastre13}.
An outcome of a sustainable assessment approach is an improved ability on the part of students to make informed judgements about the quality of their own work \citep{boud16} which here could include applicability of the tools to the problem and how to extend or change their approach while it is in progress.

Next, while allowing the students autonomy to choose the problem to address is known to be important for motivation in general \citep{urd06a}, we can expect this to be particularly important in a complex systems classroom as it addresses diversity of backgrounds in allowing students to focus on a problem from their own field. For example, Dr. Lizier suggests that for his information theory projects a biologist may examine relationships between genes, while a computer scientist may improve feature selection for machine learning; this additionally addresses motivation via authenticity of the task \citep{ashrowe14}.
The effect on motivation was obvious to the teachers here, particularly in that the students were generally found to come up with appropriate and interesting project ideas, because they are focussed on ``\textit{something they care about}'' (HS).
Prof. DeDeo was quite explicit regarding use of the project as a motivation tool, using a ``hackathon'' event as part of it, and observing that ``\textit{`A' students have been ignited by these projects.}''

Furthermore, there was clear evidence that the designs here achieved constructive alignment \citep{biggs11} quite well across learning outcomes, activities and assessment tasks.
Prof. Sayama felt that the learning outcomes in this area couldn't be assessed well with standard approaches; he was adamant that he did not want students simply memorising the material and designed the project task to assess the learning outcomes as he saw fit.
Similarly, \citet{porter18} felt that exam-based assessment used ``\textit{artificially short problems that depart substantially in both time allotted and scope from the types of problems that one actually studies in network science}'' (i.e. it was not authentic assessment).
Along these lines, each teacher explicitly articulated the learning goals associated with the major project, being (in general) to create an analysis involving intelligent application of the appropriate tools to ``\textit{tell a story}'' (SDD) about the model/data set, focussing on ``\textit{why you are doing it and how}'' (HS).
The ``story-telling'' aspect was quite important, since this was generally the focus of assessment rather than the model/analysis itself. This aligns with the thoughts on learning outcomes for these complex systems courses in general expressed in \secRef{curriculum}.
Furthermore, this aspect allows assessment of the students' understanding of the concepts they are applying, as the generally articulated goal rather than the technical aspects.
This is consistent with the importance placed by \citet{boix07a} and \citet{boix09a} on critical reflection in interdisciplinary learning, and with the learning principle proposed by \citet{jac06a} ``Encouraging Collaboration, Discussion, and Reflection''.
Indeed the presentation tasks were cited as useful for drawing out such reflection on what the students had been doing, and Dr. Lizier and Dr. Harr\'e explicitly guide the students towards focussing on such reflection in the rubrics (marking guides) supplied to the students for these projects.
Importantly also, there was a heavy emphasis on the use of formative assessment activities for scaffolding towards the major project, which was also cited as useful for getting feedback for the teacher about student capabilities.

We can see then the role of these major projects to \textit{achieve} (not only test) the learning outcomes here, in alignment with CCSF strategies \citep{trig13a}, and indeed such projects can absorb a significant amount of teaching and learning time (cited as up to 50\% by Prof. Lindgren).
The appropriateness of complex systems to be studied in this way is summarised well by \citet{jac06a}: ``\textit{complex systems phenomena are well suited to problem- and inquiry-centered learning approaches that implement constructivist models of learning and teaching}''. %
We must note however that such techniques are certainly not yet a dominant teaching paradigm, and can challenge our ideas about teaching.
For example, %
reflective concerns were raised around the ability to give good feedback, and that the amount of teaching supervision that was able to be provided at times felt minimal.
Certainly, the importance of preparing well for these projects was made evident in the interviews and in \citep{porter18} (e.g. preparing good scaffolding tasks, supporting group formation, providing clear assessment specifications, having sample data sets ready etc.).
But with such preparation done, do we as teachers need to accept that in facilitating constructivist models in this way that there is a point at which we simply must let go?
If we are going to pursue such approaches then to a large extent we do need to put faith in these models, however the counterpoint is that having and expressing such concerns is the process we must undertake as critically reflective teachers \citep{brook95}.

Continuing the theme of problem-based learning from \secRef{scholarly}, we also see these principles being incorporated into the major projects.
Prof. Lindgren described facilitating a series of in-class workshops to formulate and begin the projects for example, and Prof. DeDeo uses a ``hackathon'' style event to focus the efforts in space and time.
Dr. Lizier begins preparation by getting students to identify examples of how the tools are used in their own domains, and having students consider and discuss how the tools could be applied to provide insights on sample data sets.
The projects of course provide peer learning opportunities within groups, particularly as Prof. Sayama observes when the groups have mixed backgrounds.
In addition however we also observe important peer interactions across groups or projects being facilitated, with activities such as pitches of tasks to other groups and/or peer-feedback sessions incorporated.
Dr. Lizier and Dr. Harr\'e incorporate students' comments and feedback on other groups' presentations, both to utilise peer learning in improving the next stage of the projects, and to assess the critical thinking capacity the students have developed.
Prof. Sayama goes so far as to include peer evaluation across groups, which he has found to influence the students to explain their approaches in an accessible fashion. Not only do these provide peer-learning opportunities, but also crucial interdisciplinary interactions for this context.
The Warwick program also contained large problem-based projects, but interestingly, Mr. Finn noted that these components were not found in any of the coursework units of study; rather, students undertook two, three month long ``mini-projects'' which constituted approximately half of the required credit load. The use of such project-based units of study (often called thesis or capstone projects) has been left out of the scope of this paper, though we note that these are observed in other programs (e.g. Sydney and Chalmers) albeit not at the same weighting of the credit load. The scale of these standalone project units may be one reason that project-based assessment was not used in coursework here, as well as that 
the time length of these units of study (at 6-7 weeks) was typically shorter than others, making it more difficult to support such major projects.

Another possibility here is whether alternative assessment tasks can be offered in order to address diversity in student capabilities, as outlined in \secRef{diversity}.
This approach was originally taken at Chalmers, before restricting entry requirements, however it was found to be difficult to adapt for only a few students.
Prof. Mitchell offers an alternative of writing a paper examining three research articles on ABM.
The key to whether this is tenable and sustainable comes down to what the learning outcomes of the course are.
For Prof. Mitchell, the alternative still develops capability (learning outcomes) of understanding and evaluating use of ABMs, although this may not be suitable in other contexts (e.g. where the learning outcome is ability to \textit{develop} ABMs when such skills are required for later units).
An alternative for such contexts has been used by Dr. Lizier and Dr. Harr\'e, being a staged rubric with rudimentary ABM development attaining a pass and more complex development resulting in higher grades; while this advantages those with computational background it may be unavoidable when we require them to build these skills.

\section{Conclusion}

This study has presented and analysed findings from interviews with pioneering teachers and learners in complex systems in the higher education context.
Complex systems has only relatively recently been taught at the higher education level, and so this study fills an important gap in documenting the experiences of teachers and learners here, and how such experiences differ from other fields.

Principally we have explored experiences in the areas of curriculum design, addressing the diversity of student backgrounds, the scholarly basis of teaching, and assessment task design.
A striking level of commonality was observed in the issues expressed as well as the strategies to handle them.
In particular, the dominant issue reported was the range of technical capabilities of students, across both computer programming and mathematical analysis, due to the diversity of backgrounds of the students.
Common strategies to address this and other issues included focussing on understanding and applying key principles with technical analysis and programming implementation as a means to this end, as well as a significant focus on problem-based learning.
Furthermore, major student-led creative projects in coursework units of study were widely used, with very common structures and learning goals, and were utilised for both achieving as well as assessing learning outcomes.

Perhaps the commonalities in our experiences should not be so surprising, since we are all part of a common `learning culture' in our discipline \citep{james14a}.
Indeed, in the contemporary environment with frequent interactions at conferences and on social media, sharing ideas and approaches in smaller fields such as complex systems -- whether explicitly or implicitly -- is facilitated more easily than in the past.

Finally, we note the significant extent to which the teaching approaches described here already align with known effective practice in the educational literature, despite the relatively small exposure many educators here have had to formal education theory.
We hope that our discussion of such experiences with reference to the education literature can bring wider exposure to such theory in the teaching of complex systems, and prompt more extensive investigation and discussion on how to increase effectiveness of teaching and learning complex systems in higher education.

\section*{Author contributions}

JL conceived the study. JL, MH and AL conducted background research. JL interviewed MM, SDD, KL, CF and HS, and performed thematic analysis to compare and contrast the responses. JL drafted the manuscript, JL and AL situated the experiences with respect to the educational literature, and all authors edited and approved the manuscript.

\section*{Data availability}

The interviews analysed during this study are not publicly available because the interviewees are directly identifiable, and the conversations contain material (beyond that which publication consent has been granted for here) around which there are privacy and commercial confidentiality concerns.

\section*{Conflicts of interest}

The authors declare no conflicts of interest.

\section*{Funding Statement}

JL was supported through the Australian Research Council DECRA grant DE160100630.

\section*{Acknowledgements}

JL thanks Dr. Amani Bell for helpful feedback on the interview design and early drafts of the manuscript, and JL and MH thank Dr. Graham Hendry for comments on background research.

\bibliography{references}

\end{document}